\documentclass[prb,showpacs,twocolumn,floatfix,superscriptaddress]{revtex4}

\usepackage{epsfig}
\usepackage{amsmath,amssymb,mathrsfs}
\def\up{\uparrow}
\def\down{\downarrow}

\begin{document}

\title{Strongly modulated transmission of a spin-split quantum wire with local Rashba interaction}
\author{David S\'anchez}
\affiliation{Departament de F\'{\i}sica, Universitat de les Illes Balears,
E-07122  Palma de Mallorca, Spain}
\author{Lloren\c{c} Serra}
\affiliation{Departament de F\'{\i}sica, Universitat de les Illes Balears,
E-07122  Palma de Mallorca, Spain}
\affiliation{Institut Mediterrani d'Estudis Avan\c{c}ats IMEDEA (CSIC-UIB), 
E-07122 Palma de Mallorca, Spain}
\author{Mahn-Soo Choi}
\affiliation{Department of Physics, Korea University,
Seoul 136-701 Korea}

\date{\today}

\begin{abstract}
We investigate the transport properties of ballistic quantum wires
in the presence of Zeeman spin splittings
and a spatially inhomogeneous Rashba interaction.
The Zeeman interaction is extended along the wire and produces gaps in the energy
spectrum which allow electron propagation only for spinors lying along
a certain direction. For spins in the opposite direction the waves are evanescent
far away from the Rashba region, which plays the role of the scattering center.
The most interesting case occurs when the magnetic field is perpendicular
to the Rashba field. Then, the spins of the asymptotic wavefunctions are
not eigenfunctions of the Rashba Hamiltonian and the resulting coupling
between spins in the Rashba region gives rise to sudden changes of the
transmission probability when the Fermi energy is swept along the gap.
After briefly examining the energy spectrum and eigenfunctions of a wire
with extended Rashba coupling, we analyze the transmission through a
region of localized Rashba interaction, in which a double interface
separates a region of constant Rashba interaction from wire leads
free from spin-orbit coupling. 
For energies slightly above the propagation threshold,
we find the ubiquitous occurrence of transmission zeros (antiresonances)
which are analyzed by matching methods in the one-dimensional limit.
We find that a a minimal tight-binding model yields
analytical transmission lineshapes of Fano antiresonance type.
More general angular dependences of the external magnetic field
is treated within projected Schr\"odinger equations with Hamiltonian
matrix elements mixing wavefunction components. Finally,
we consider a realistic quantum wire where the energy subbands
are coupled via the Rashba intersubband coupling term and discuss
its effect on the transmission zeros.
We find that the antiresonances are robust against intersubband mixing,
magnetic field changes,
and smooth variations of the wire interfaces, which paves the way
for possible applications of spin-split Rashba wires as spintronic
current modulators.
\end{abstract}

\maketitle

\section{INTRODUCTION}

\subsection{Motivation}
Since the pioneering Datta-Das proposal of an electronic field-effect
transistor in which the current flow is controlled by magnetic means only,\cite{dat90}
the study of the Rashba\cite{ras60,byc84} spin-orbit interaction in one-dimensional (1D)
and quasi-one-dimensional
ballistic channels (quantum wires) has attracted
a lot of interest.\cite{sat99,mor99,mir01,kis01,mol01,egu02,gov02,fev02,bul02,and03,stre03,scha04,per04,nes04,cah04,wan04,she04,zha05,kno05,per05,rom05,deb05,ser05,zhan05,zhan06,rey06,san06,jeo06,sha06,zha06,per07,nev07}
Precise tunability of the strength of the Rashba coupling
has been also experimentally demonstrated in quantum
wells.\cite{nit97,eng97,gru00}
Typically, semiconductor quantum wires are built from
two-dimensional electron gases formed at the interface
of a semiconductor heterostructure when the lateral motion
of electrons is restricted by a transversal confinement potential
to effective widths of the order of the de Broglie electron
wavelength. For very clean quantum wires (e.g., quantum point
contacts) transport is ballistic and conductance is quantized
to integer values of $e^2/h$.\cite{wee88,wha88}

The presence of impurities or defects in the vicinity of the constriction
destroys conductance quantization.\cite{chu89,bag90,fai90,tek91,gur93,noc94}
A striking effect arises when the impurity potential
is attractive and enables
the existence of at least one bound state whose energy is
degenerate with the continuum band of propagating states.
As a consequence, for energies close to the transition threshold 
a direct transmission channel can interfere with a
wave trajectory that travels across the bound state
and this interference is destructive, leading
to enhanced backscattering and Fano asymmetric lineshapes.\cite{fano,cer73,gor00,kob02}
Recently, two of us\cite{san06} have demonstrated that a spin-orbit
interaction of the Rashba type localized in an infinitely long
quantum wire plays a role similar to an attractive potential and pronounced dips are seen
in numerical simulations of the conductance curves.\cite{she04,zhan05,san06,zha06}
It is remarkable that the Rashba interaction
provides both the attractive potential that supports bound states\cite{val04,cse04}
and the mixing term
that couples the localized and the propagating states.\cite{san06}
Interestingly, when charging effects are taken into account,
Coulomb blockade resonances can be tuned directly modulating the strength
of the Rashba coupling.\cite{lop07}

A magnetic field applied in the wire plane leads to
Zeeman spin splitting of the 1D modes. Evidence of this
is shown in the appearance of conductance plateaus
at odd multiples of $e^2/h$.\cite{wha88} In the first plateau the current
is fully polarized since only one spin species is allowed to propagate.
Quantum states with opposite spin are evanescent asymptotically
and do not take part in electron transport unless there exist
inhomogeneities that give rise to resonances or Fano-type
interferences in which case evanescent states are crucial.
Reference \onlinecite{ser07} presents a theoretical method to calculate evanescent
states in quantum wires with uniform Rashba interaction.

To determine the full transmission pattern of a generic quantum wire, one must
first analyze the energy spectrum of the wire. For quantum wires with
uniform Rashba interaction in the absence of external magnetic
fields, free-electron energy bands are parabolas shifted
apart for opposite spin directions.\cite{mol01} The splitting size
is proportional to the spin-orbit interaction strength $\alpha$
and in the quasi-1D case the Rashba interaction produces
anticrossings between bands corresponding to opposite spins
and adjacent modes.\cite{mor99,gov02} Moreover, the propagation threshold
is shifted, compared to the case with no spin-orbit coupling,
down to an energy $m\alpha^2/2\hbar^2$. In the presence of an in-plane magnetic field,
the energy spectrum changes dramatically even for arbitrarily
small fields. The field can be either externally applied or originated
from stray fields of the ferromagnets coupled to the wire in the Datta-Das
setup.\cite{dat90} It is shown\cite{stre03,per04,ser05}
that the interplay between the magnetic field
and the Rashba interaction leads to the openings of gaps in the 1D
energy bands at small wavenumbers.
In a quasi-1D wire most of the energy dispersions around the gap
form energy minima locally in contrast to the maxima
encountered in the 1D case.\cite{ser05}
In those energy windows in which the gap consists of an energy
local maximum followed by a local minimum the conductance curves
present anomalous steps for chemical potentials within the gap.\cite{per04,ser05}

In short, external magnetic fields lead to the formation of
energy gaps in the spectrum while local Rashba interactions
produce Fano-type antiresonances due to the formation of quasi-bound
states coupled to the channel of direct transmission states.
Therefore, we expect a rich interplay between in-plane fields and
localized Rashba spin-orbit couplings
in the transport properties of a ballistic quantum wire.
This paper presents a generic theoretical description
of the quantum transmission of an electron subject to
Zeeman splittings and spatially modulated Rashba fields.

\subsection{Main findings}

We find the occurrence of {\em exact} tranmission zeros in the conductance
curves of a Zeeman-split wire with local Rashba interaction as a function of the Fermi
energy. Central to the existence of the transmission zeros are the
formation of a Zeeman gap arising from an in-plane magnetic field
and the role of the evanescent states
within the Rashba region. In fact, the Rashba interaction couples
the propagating and evanescent states precisely in the interior
of the Rashba region. The transmission antiresonances
are almost universal, showing a Fermi energy with
vanishingly small transmission at moderately low magnetic fields.
This might be relevant for applications
since it provides two operation points for working transistors (low and high
current states). It is important to stress that these transmission zeros
are fundamentally distinct from the suppressed transmission that
may take place in a Datta-Das setup due to spin precession\cite{dat90}
even in the presence of in-plane magnetic fields.\cite{nev07}
The antiresonance position can be tuned with a slight change of $\alpha$
and are robust against changes of the magnetic field.
We only require that the Fermi energy lies within the gap.

\subsection{Outline}

The outline of the paper is as follows. Section~\ref{sec-1d}
is devoted to analyzing the transport properties of a 1D wire,
where only one subband is taken into account and Rashba-induced
intersubband coupling is then neglected. In Sec.~\ref{sec-1d-ext}
we discuss the eigenstates and energy spectrum of a 1D wire
subject to Rashba interaction and Zeeman spin splittings.
In Sec.~\ref{sec-1d-di} we consider
a finite Rashba region with constant Rashba strength
and a magnetic field pointing in a direction perpendicular to the Rashba field
and calculate the transmission within
the scattering formalism and numerical matching.
A tight binding description of the problem
is considered in Sec.~\ref{sec-1d-tb}. We derive an exact expression
for the transmission in the limit of a minimal Rashba region
and discuss the Fano form of the lineshape. To end Sec.~\ref{sec-1d} we 
present in Sec.~\ref{sec-1d-ang} results for the angular dependence
of the magnetic field direction.
In Sec.~\ref{sec-q1d} we examine a quasi-1D wire. The numerical results
are in agreement with the 1D case, thus demonstrating that the sharp antiresonances
are robust even when intersubband coupling is present like in realistic wires. 
We also discuss the case of an arbitrary dependence
of the Rashba strength with the position and compare our results
when the Rashba interaction smoothly increases at the interfaces.
Finally, Sec.~\ref{sec-concl} contains the conclusions.

\section{ONE-DIMENSIONAL WIRE}\label{sec-1d}

We consider a 2D electron gas formed in the $x$-$y$ plane
due to strong confinement in the $z$ direction. As a result
of the interfacial electric field, there arises a spin-orbit
coupling of the Rashba type with a Hamiltonian given by
\begin{equation}
\label{Eq_Rash}
{\cal H}_R = \frac{\alpha(x)}{2\hbar}
(p_y \sigma_x - p_x \sigma_y) +
{\rm H.c.}\; ,
\end{equation}
where the Rashba strength $\alpha(x)$ can be spatially modulated.
The limit of a purely 1D system is obtained by further
constraining the electron motion along, e.g., the $x$ direction. 
Then, the $p_y\sigma_x$ term in Eq.~(\ref{Eq_Rash}) is neglected
and the Rashba interaction plays the effective role
of a momentum-dependent magnetic field with a direction along
the $y$ axis. In the following section we discuss
the spectrum and eigenfunctions of a 1D wire when the Rashba
strength is uniform and the external magnetic field points
either in the $x$-$y$ plane (``in-plane'' field)
or along the vertical $z$ direction (``perpendicular'' field).

\subsection{Extended Rashba interaction}\label{sec-1d-ext}

\subsubsection{In-plane field}
We consider an in-plane magnetic
field with arbitrary direction, $\vec{B}=(B\cos\theta,B\sin\theta,0)$,
giving rise to Zeeman interaction.\cite{per04,ser05}
Then, the single-particle Hamiltonian reads,
\begin{eqnarray}\label{ham1d}
{\cal H}_{{\rm 1D}}&=&\frac{-\hbar^2}{2m}\frac{d^2}{dx^2}+
\frac{\Delta_Z}{2}  (\sigma_x\cos\theta+\sigma_y\sin\theta) \nonumber \\
&-&\frac{1}{2\hbar}\sigma_y\{\alpha(x), -i\hbar\frac{d}{dx}\}
\; ,
\end{eqnarray}
where $\Delta_Z=g \mu_B B$ is the Zeeman splitting.
When we assume a constant Rashba strength
$\alpha(x)\equiv\alpha={\rm const.}$ from $-\infty$ to $\infty$,
the Hamiltonian ${\cal H}_{{\rm 1D}}$ is diagonalized
using the spinor wavefunctions (we take the spin quantization direction
along $z$),
\begin{equation}\label{wf1d}
\psi_\eta(x)= \frac{1}{\sqrt{2}}\left( \begin{array}{ccc}
e^{i\Omega_k/2} \\
\eta e^{-i\Omega_k/2}\end{array} \right) e^{i k x}
\; ,
\end{equation}
where $\eta=\pm$ is the branch-splitting quantum index and
$k$ is the wavevector associated to free motion along $x$. The
spin orientation is determined from
\begin{equation}\label{omegak}
\tan \Omega_k(\theta) = \frac{\alpha k-(\Delta_Z/2)\sin\theta}
{(\Delta_Z/2)\cos\theta}
\; .
\end{equation}
We note that there exists no common spin quantization axis since
$\Omega_k$ depends on $k$.\cite{ser05} This is due to the existence of a magnetic
field since for $\Delta_Z=0$ the spinors lie along $y$ (the Rashba axis).
The effect is akin to a 2D system with Rashba interaction for which
$\tan\Omega=k_x/k_y$. However, in the 2D case the spin orientation
is always perpendicular to $\vec k$ whereas in the 1D case only for
asymptotically large momenta ($|k|\to\infty$) the spin is quantized
along $y$. A similar effect arises from Rashba intersubband coupling
in quasi-1D systems.\cite{gov02}
\begin{figure}
\centerline{
\epsfig{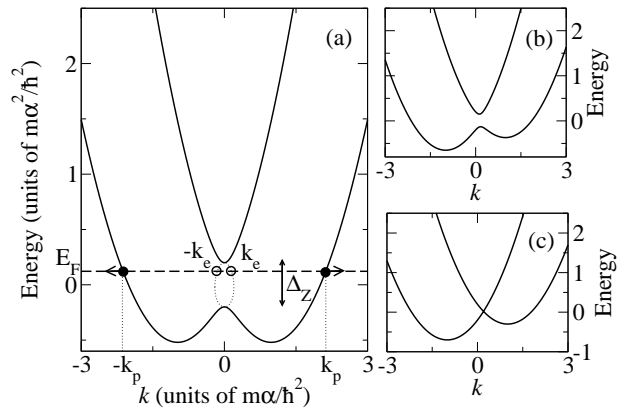}
}
\caption{Energy spectrum for a 1D quantum wire with uniform Rashba coupling
for a Zeeman splitting $\Delta_Z=0.4m\alpha^2/\hbar^2$ and different magnetic field
angles: (a) $\theta=0$, (b) $\theta=\pi/4$ and (c) $\theta=\pi/2$.}
\label{fig1}
\end{figure}

From the Schr\"odinger equation, ${\cal H}_{{\rm 1D}}\psi_\eta=E_\eta\psi_\eta$,
one finds the energy spectrum,
\begin{equation}\label{en1d}
E_\eta(k)=\frac{\hbar^2 k^2}{2m}+\eta\sqrt{\alpha^2k^2+
(\Delta_Z/2)^2-\alpha k\Delta_Z\sin\theta}
\; .
\end{equation}
For $\theta=0$ the case of interest occurs for small Zeeman splittings,
$\Delta_Z<2m\alpha^2/\hbar^2$, see Fig. \ref{fig1}(a).
The lowest branch of the spectrum develops a local maximum around
$k=0$ which has important consequences for transport.\cite{stre03,per04,nes04,ser05}
As a result, there arises a pseudogap between the two spectrum branches.
For Fermi energies lying in the pseudogap region, $-\Delta_Z/2<E_F<\Delta_Z/2$
there is only a wavefunction with a given spin direction for
each mover (right-moving or left-moving).
In addition, there also exist \emph{evanescent} waves which are crucial
when the Rashba interaction is confined to a finite region.
Outside the pseudogap region, there are four real wavevectors
for a given $E_F$.

Increasing $\theta$ from $0$ to $\pi/2$ leads to the progressive reduction
of the pseudogap size, see Fig. \ref{fig1}(b) and \ref{fig1}(c).
For $\theta=\pi/2$ the gap
vanishes and the spinors point along the $y$ axis since in this case
the field axis and the Rashba axis coincide.

We note in passing that when $\alpha(x)$ is constant
(or, more generically, an even function of $x$),
there exists a symmetry property of
the Hamiltonian given by Eq.~(\ref{ham1d}). Let us concentrate
on the case $\theta=0$. Thus, ${\cal H}_{\rm 1D}$
is invariant under the transformation
\begin{equation}
\label{RashbaRabi::eq:9}
\hat Z
= i\Pi\exp\left[-i\pi\hat\sigma_x\right]
= \Pi\hat\sigma_x
\,,
\end{equation}
namely, the rotation by $\pi$ around $x$ axis in the spin space
followed by the parity operator $\Pi$, which yields inversion in the
$x$ direction.  The additional
factor of $i$ is to ensure that $\hat Z^2=1$. Similar symmetry
properties have been discussed in Refs.\onlinecite{bul99,kis01,deb05}
which find that the
{\em spin parity}, i.e., the combination of parity and a Pauli matrix,
is a constant of motion for $B=0$. Here, $\Pi\hat\sigma_x$ commutes
with ${\cal H}_{\rm 1D}$ even for nonzero fields when $\theta=0$.
Hence, one can find a common basis of eigenstates
for ${\cal H}_{\rm 1D}$ and $\hat Z$.
The wave functions given by Eq.~(\ref{wf1d})
are not eigenstates of $\hat Z$. In fact,
$|\psi_{\eta,+|k|}\rangle=\eta\hat Z|\psi_{\eta,-|k|}\rangle$
since $\Omega\to-\Omega$ when $|k|\to-|k|$. Therefore,
one could construct states with definite parities with regard to $\hat Z$
from even ($e$) and odd ($o$) combinations of $|\psi_{\eta,\pm|k|}\rangle$, i.e.,
$|\psi_\eta^{e/o}\rangle=|\psi_{\eta,+|k|}\rangle\pm\eta|\psi_{\eta,-|k|}\rangle$.

\subsubsection{Perpendicular field}
For fields pointing along $z$, $\vec{B}=(0,0,B)$ the Zeeman term
in Eq.~(\ref{ham1d}) is expressed as $(\Delta_Z/2)\sigma_z$. The spectrum
is identical to the $\theta=0$ case, Fig. \ref{fig1}(a). Therefore, for the
sake of the present discussion, the cases $\vec{B}$ parallel to $z$ and
$\vec{B}$ parallel to $x$ are equivalent in the 1D case
since both are perpendicular to the Rashba field direction (along $y$).
Of course, in the quasi-1D
case one should also take into account orbital effects but in this
section we can neglect this. Let us focus on the $\eta=-$ branch
and energies within the pseudogap. We will find that strong transmission
changes take place in that region.

For $-\Delta_Z/2<E_F<\Delta_Z/2$  there are two propagating solutions
\begin{equation}\label{eq_psip}
\psi(x)= \left( \begin{array}{ccc}
\pm\sin{\Omega_p\over 2} \\
i\cos{\Omega_p\over 2}\end{array} \right) e^{\pm i k_p x}
\,,
\end{equation}
with wavevector [see Fig. \ref{fig1}(a)]
\begin{equation}\label{eq_kp}
k_p=\sqrt{k_F^2+2k_R^2+\sqrt{k_B^4+4k_R^2(k_R^2+k_F^2)}}
\,,
\end{equation}
where we have defined
\begin{eqnarray}
k_F&=&\sqrt{2mE_F/\hbar^2}\,, \\
k_R&=&m\alpha/\hbar^2\,, \\
k_B&=&\sqrt{m\Delta_Z/\hbar^2}
\,.
\end{eqnarray}
As discussed above, the angle $\Omega_p$ depends on the wavevector:
\begin{equation}
\tan{\Omega_p}=\frac{2 k_R k_p}{k_B^2} \,.
\end{equation}

Further, the pseudogap region admits two more solutions which are
evanescent waves. For $E_F<-\hbar^2 B^2/2m\alpha^2$ (i.e.,
$k_F^2<k_B^4/4 k_R^2$) we find,
\begin{equation}\label{eq_psie}
\psi(x)= \left( \begin{array}{ccc}
\mp\sinh{\Omega_e\over 2} \\
\cosh{\Omega_e\over 2}\end{array} \right) e^{\pm k_e x}
\,,
\end{equation}
For $E_F>-\hbar^2 B^2/2m\alpha^2$
one must make the replacements $\sinh \to\cosh$
and $\cosh \to\sinh$. Moreover,
\begin{equation}\label{eq_ke}
k_e=\sqrt{-k_F^2-2k_R^2+\sqrt{k_B^4+4k_R^2(k_R^2+k_F^2)}}
\,.
\end{equation}
and 
\begin{equation}
\tanh{\Omega_e}=\frac{2 k_R k_e}{k_B^2} \,.
\end{equation}
In Fig. \ref{fig1}(a) we illustrate the ``dispersion'' relation
for the evanescent states and the location of $\pm k_e$.
The evanescent states make sense only for nonvanishing $\Delta_Z$
as can be readily seen by setting $k_B=0$ in Eqs.~(\ref{eq_kp})
and~(\ref{eq_ke}). $k_p$ becomes $k_R+\sqrt{k_R^2+k_F^2}$ and
$k_e$ becomes pure imaginary, $k_e=i(k_R-\sqrt{k_R^2+k_F^2})$.
This corresponds to four propagating solutions (two left-moving
and two right-moving) with a definite spin direction.\cite{mol01}

\subsection{Local Rashba interaction}\label{sec-1d-di}
\begin{figure}
\centerline{
\epsfig{file=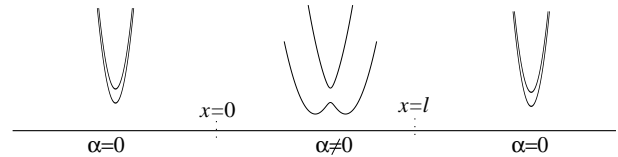,angle=0,width=0.45\textwidth,clip}
}
\caption{Schematic representation of a local Rashba interaction in a 1D
quantum wire.}
\label{fig4}
\end{figure}

We now consider a double interface at $x=0$ and $x=l$ between a normal conduction band
($x<0$ and $x>l$) and a region of localized Rashba interaction
extending from $x=0$ to $x=l$, see Fig. \ref{fig4}.
Then, $\alpha(x)\equiv\alpha={\rm const.}$ for $0<x<l$ and $\alpha(x)=0$ elsewhere.
Our basic goal is to find the transmission $T$
through the Rashba region when the magnetic field is present all along the wire,
producing a Zeeman gap.
For convenience, we take the $\vec B$ direction along $z$ since
the solutions for $x<0$ are simpler to write down.

We are interested in energies inside the spin pseudogap,
$-\Delta_Z/2<E_F<\Delta_Z/2$, for which a spin-down (spin-up) electron wave
is propagating (evanescent). In the scattering problem, an electron with spin down
is injected from $-\infty$ and reflected with a certain probability
$R=1-T$. Since the spin quantization axis in the region $0<x<l$ depends on the
wavevector, we must also take into account the spin-up evanescent waves
at $x<0$ and $x>l$. As a consequence, the scattering wave function
for $-\Delta_Z/2<E_F<-\hbar^2 B^2/2m\alpha^2$ reads,
\begin{widetext}
\begin{equation} \label{eq_psix}
\psi(x)=\left\{ \begin{array}{l} \left( \begin{array}{l}
0 \\
1\end{array} \right) e^{i k_1 x}+
\left( \begin{array}{l}
0 \\
1\end{array} \right) A e^{-i k_1 x}+
\left( \begin{array}{l}
1 \\
0\end{array} \right) B e^{k_2 x} \,\,\,\,\,\, x<0 \\
\left( \begin{array}{l}
\sin{\Omega_p\over 2} \\
i \cos{\Omega_p\over 2} \end{array} \right) C e^{i k_p x}
+ \left( \begin{array}{l}
-\sin{\Omega_p\over 2} \\
i \cos{\Omega_p\over 2} \end{array} \right) D e^{-i k_p x} \\
+\left( \begin{array}{l}
\sinh{\Omega_e\over 2} \\
\cosh{\Omega_e\over 2} \end{array} \right) F e^{-k_e x}
+\left( \begin{array}{l}
-\sinh{\Omega_e\over 2} \\
\cosh{\Omega_e\over 2} \end{array} \right) G e^{k_e x}
\,\,\,\,\,\, 0<x<l \\
\left( \begin{array}{l}
0 \\
1\end{array} \right) J e^{i k_1 (x-l)}+
\left( \begin{array}{l}
1 \\
0\end{array} \right) K e^{-k_2 (x-l)} \,\,\,\,\,\, x>l
\end{array}\right.
\,,
\end{equation}
\end{widetext}
whereas for $-\hbar^2 B^2/2m\alpha^2<E_F<\Delta_Z/2$ one should make
the replacements $\sinh\to\cosh$ and $\cosh\to\sinh$ in the expressions for the
evanescent states of the Rashba region.
In Eq.~(\ref{eq_psix})
the wavevector $k_1=\sqrt{k_F^2+k_B^2}$ is written in terms of
the Fermi wavevector $k_F^2=2mE_F/\hbar^2$ and $k_B^2=m \Delta_Z/\hbar^2$
defined as before. We note that $k_F$ can be pure imaginary for $E_F<0$, though
the physical wavevector, $k_1$, is always a quantity manifestly positive and real.
The evanescent wave is described with an exponentially decreasing
wave with range $\sim 1/k_2$ where $k_2=\sqrt{k_B^2-k_F^2}$,
thereby the probability of finding a spin-up on the left or right sides
is nonzero. At small $\Delta_Z$ the propagating states with wavevector $\pm k_p$
has its spin approximately along $\pm y$ [$\Omega\approx\pi/2$ in Eq.~(\ref{wf1d})].

We numerically find the coefficients $A$, $B$, $C$, $D$, $F$, $G$, $J$ and $K$
from matching equations.
At the interfaces
the wavefunction must be continuous, e.g., $\psi(0^-)=\psi(0^+)$.
Moreover, the flux, given by the velocity operator $\hat v$,
must also be continuous, i.e., $\hat v\psi(0^-)=\hat v\psi(0^+)$.
On the normal sides, $\hat v$ is
trivially given by $-i(\hbar/m)d/dx$ while in the spin-orbit region
one has\cite{mol01}
\begin{equation}
\hat v=\frac{i\hbar}{m}\left( \begin{array}{ccc}
-\frac{d}{dx} & k_R\\
-k_R & -\frac{d}{dx}\end{array} \right)
\,.
\end{equation}
Importantly, the flux associated to Eq.~(\ref{eq_psip}),
which is proportional to
\begin{equation}\label{eq_fluxp}
\langle\hat v\rangle=\pm k_p-k_R \sin\Omega_p\,,
\end{equation}
is positive (negative) for $+k_p$ ($-k_p$).

\begin{figure}
\centerline{
\epsfig{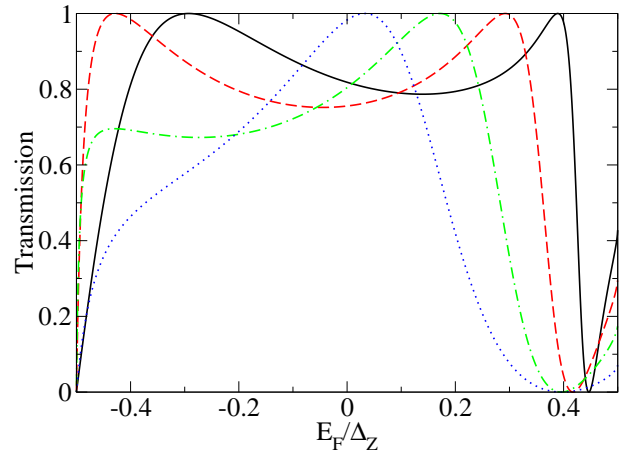}
}
\caption{(Color online) Transmission through a Rashba region
for $\Delta_Z=12.8 \hbar^2/ml^2$ and different spin-orbit intensities:
$\alpha=6.4 \hbar^2/ml$ (full line), 6.64 (dashed), 6.88 (dot-dashed), 7.12 (dotted).}
\label{fig5}
\end{figure}

The transmission and the reflection are simply given by $T=|J|^2$
and $R=|A|^2$, respectively.
Since we are interested in the relative influence
of $\alpha$ and $\Delta_Z$ while keeping the Rashba region length $l$ constant,
in the numerical results we give the energies as a function of the energy unit
$\hbar^2/m l^2$. A characteristic transmission curve is plotted in Fig. \ref{fig5} for
$\Delta_Z=12.8 \hbar^2/ml^2$ whereas $\alpha$ is slightly varied.
We make the important observation that
there arises an exact transmission zero near $E_F=\Delta_Z/2$ when $\alpha=6.4$.
On increasing $\alpha$, the resonance position shifts to lower energies
and, at the same time, the resonance broadening is enhanced.
This dependence with $\alpha$ will become clear later, when
we discuss the tight-binding model.

\begin{figure}
\centerline{
\epsfig{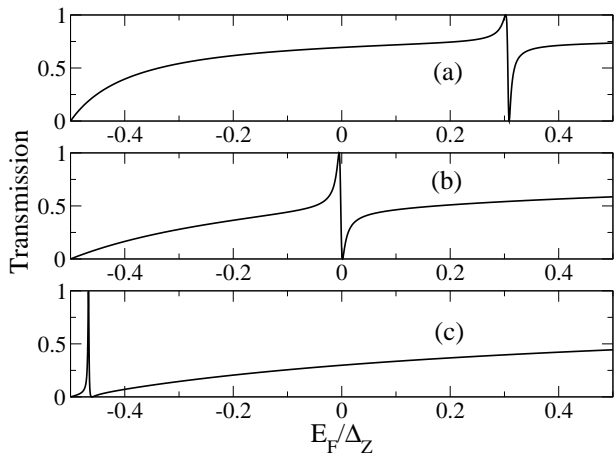}
}
\caption{Transmission versus Fermi energy for $\alpha=6.4 \hbar^2/ml$
and different Zeeman splittings: $\Delta_Z=0.384 \hbar^2/ml^2$ (a), 0.128 (b) and 0.064 (c).}
\label{fig6}
\end{figure}

The transmission for $\alpha=6.4 \hbar^2/ml$ and various magnetic fields
is shown in Fig. \ref{fig6}.
The transmission curves are reminiscent of the Fano lineshapes.
An interesting question is thus whether the transmission behavior
is indeed related to a Fano-type interference effect.
While in strict one-dimensional
systems the interference giving rise to Fano lineshapes is not possible due
to the existence of one channel only, in this case and due
to the Zeeman splitting there exist two modes, namely, the propagating mode
(spin down) and the evanescent (spin up) mode. Both modes become coupled locally
within the Rashba region. As a result, the effect is due to a 
subtle combination of spin-orbit interaction and Zeeman splitting
which leads to destructive interference in the Rashba region.
A simplified model, discussed below, will shed light on this.
For the moment, we note that Fig. \ref{fig6} shows that it is sufficient
to have a rather small Zeeman gap $\Delta_Z^{\rm crit}$ above which
the antiresonance develops.
For $\alpha=6.4 \hbar^2/ml$ we find $\Delta_Z^{\rm crit}\simeq 0.064 \hbar^2/ml^2$.

The energy and length scales we considered above are within the scope of present
techniques. E.g., for a Rashba region of size $l=2$~$\mu$m the value
$\alpha=6.4 \hbar^2/ml$ corresponds to $\alpha\simeq 10$~meV~nm, which
is accessible in an InAs wire.\cite{nit97} The Zeeman energy
$\Delta_Z=12.8 \hbar^2/ml^2$ used in Fig.~\ref{fig5}
corresponds to a magnetic field $B\simeq 10$~mT
in the same material and $\Delta_Z^{\rm crit}$ in Fig.~\ref{fig6}(c) is only
60$\mu$T. Notably, the effect scales with $l$. Therefore, a smaller $\alpha$
would require a larger wire for the antiresonance to be observable.

\subsection{Tight-binding model}\label{sec-1d-tb}
\begin{figure}[t]
\centerline{
\epsfig{file=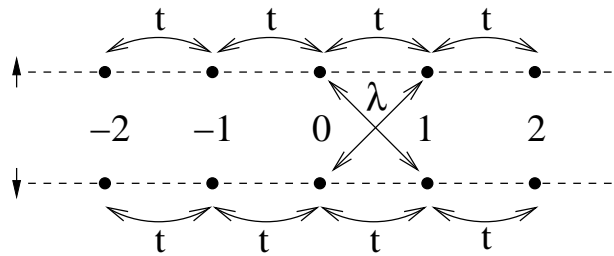,angle=0,width=0.45\textwidth,clip}
}
\caption{In the minimal tight-binding model,
the system consists of a linear chain of
coupled sites with a localized Rashba interaction
at sites $n=0$ and $n=1$.}
\label{figtb}
\end{figure}

The continuum model discussed above leads to remarkable predictions for
the conductance of a spin-split quantum wire with a local Rashba interaction
but to gain further insight it would be highly desirable to have
a simplified model capable of yielding closed analytical formulas
for the dip position and shape. In this subsection we consider
a discretized version of the Hamiltonian ${\cal H}_{\rm 1D}$ [Eq.~(\ref{ham1d})].
The infinite 1D wire is modeled with a linear chain of sites.
Thus, we obtain the following tight-binding Hamiltonian,\cite{and89}
\begin{eqnarray}
\label{eq_h2}
{\cal H}_{\rm tb}&=&\sum_i \varepsilon_{\sigma} c_{i\sigma}^\dagger c_{i\sigma}
-\sum_{\langle ij\rangle} t (c_{i\sigma}^\dagger c_{j\sigma}+{\rm H.c.}) \nonumber \\
&+&\lambda \sum_n (c_{n\uparrow}^\dagger c_{(n+1)\downarrow}
-c_{n\downarrow}^\dagger c_{(n+1)\uparrow}+{\rm H.c.})\,,
\end{eqnarray}
where the summations over $i$ and $j$ are carried out on an infinitely
extended 1D wire and the last summation is over the sites of the Rashba
region. In this equation, $t=\hbar^2/2ma^2$ couples nearest neighbors,
$\lambda=\alpha/2a$ is the Rashba interaction strength which
couples electronic states with opposite spin directions along $z$
$(\sigma=\{\up,\down\})$, and $a$ is the lattice parameter.
The on-site energies are given by $\varepsilon_{\sigma}=s\Delta_Z/2$
[$s=+(-)$ for $\sigma=\up(\down)$].
The Hamiltonian ${\cal H}_{\rm tb}$ is equivalent,
in the limit $a \to 0$, to ${\cal H}_{\rm 1D}$ whose transport properties
have been analyzed above. Here, in order to obtain
simplified expressions we consider a localized
Rashba interaction restricted only to two sites,
0 and 1 (see Fig. \ref{figtb}). We note that this is the minimal
model that characterizes the transport properties
of a 1D wire with a Rashba region.

\begin{figure}[t]
\centerline{
\epsfig{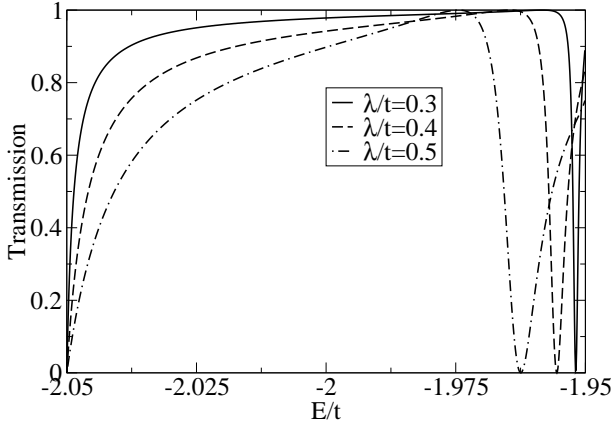}
}
\caption{Transmission versus Fermi energy for $\Delta_Z=0.1$
and different spin-orbit strengths.
All energies are given in units of $t$.
}
\label{figtb2}
\end{figure}

For $\Delta_Z=0$ and $\lambda=0$ the energy band spectrum is
given by the well known expression $E=-2t\cos ka$. In the presence
of a magnetic field, the spectrum becomes spin split.
We now focus on an energy range
close to the band bottom, $-2t-\Delta_Z/2\le E \le -2t+ \Delta_Z/2$.
Then, the eigenfunctions corresponding to spin $\down$ ($\up$)
are propagating (evanescent) waves.
Since we intend
to solve the scattering problem of an $\down$-electron wave
impinging onto the Rashba region, we now introduce
the wave amplitudes
$\psi_{n\downarrow}=e^{i k_\downarrow n a}+re^{-i k_\downarrow n a}$ for $n\le -1$,
and $\psi_{n\downarrow}=\tau e^{i k_\downarrow n a}$ for $n\ge 2$,
with $\tau$ and $r$ the transmission and reflection probability amplitudes.
The wavenumber $k_\downarrow$ is related
to the total energy $E$ by means of
$k_\downarrow a=\cos^{-1} [(E+\Delta_Z/2)/(-2t)]$
For electrons with spin $\up$, their energy outside the Rashba region
falls below the band bottom. As a result, we take the wave amplitudes
given by $\psi_{n\uparrow}=c e^{k_\uparrow n a}$ for $n\le 0$ and
$\psi_{n\uparrow}=d e^{-k_\uparrow n a}$ for $n\ge 1$,
where $k_\uparrow a=\cosh^{-1} [(E-\Delta_Z/2)/(-2t)]$
Substituting the total wavefunction into the Schr\"odinger equation
and projecting over the sites of the Rashba region $\langle n \sigma|$,
we find the transmission,
\begin{equation}\label{eq_tau}
\tau=\left[1+\frac{\lambda^2}{t^2}
\frac{t^2(1-e^{k_\uparrow a+ik_\downarrow a})+(\lambda^2+t^2) e^{2ik_\downarrow a}}
{(\lambda^2-t^2(e^{2k_\uparrow a}-1))(e^{2ik_\downarrow a}-1)}\right]^{-1}
\,,
\end{equation}
which allows to determine the {\em exact} condition for the occurrence
of zeros in the transmission function $T=|\tau|^2$:
\begin{equation}\label{eq_e0}
E^{(0)}=\frac{\Delta_Z}{2}-2t\frac{1+\lambda^2/2t^2}{\sqrt{1+\lambda^2/t^2}}
\,.
\end{equation}
This expression has a very appealing form.
For a given value of $\lambda$ the antiresonance energy $E^{(0)}$
lies to the left of $-2t+\frac{\Delta_Z}{2}$, as shown in the numerical
results of Figs.~\ref{fig5} and~\ref{figtb}.
Moreover, it predicts that the dip shifts to lower
energies as the Rashba interaction strength increases. This is
reproduced in Fig. \ref{figtb2} where we plot a characteristic
$T$ as a function of energy for $\Delta_Z=0.1t$.
In addition, the dip broadens as $\lambda$ increases, in excellent agreement
with the numerical results of the continuum model, see Fig.~\ref{fig5}.
Equation~(\ref{eq_e0}) also explains why the critical Zeeman splitting
$\Delta_z^{\rm crit}$ below which the dip disappears is so small
since the antiresonance is observable only if $E^{(0)}$ lies above
the band bottom, i.e., $E^{(0)}\ge -2t-\frac{\Delta_Z}{2}$.
It then follows that $\Delta_z^{\rm crit}=2t f(\lambda/t)$ where
$f(x)=(1+x^2/2)/\sqrt{1+x^2}-1$ is a slowly increasing function of $x$ for
$0\le x\le 1$. As a result, $\Delta_z^{\rm crit}\lesssim 0.12 t$, which
is at least two orders of magnitude smaller than the bandwidth.
Incidentally, we also find an upper bound of the Rashba strength,
$\lambda^{\rm crit}=\sqrt{\Delta_Z^2/2+2\Delta_Z t +
(t+\Delta_Z/2)\sqrt{\Delta_Z^2+4t\Delta_Z}}$, above which the dip vanishes. 
For $\Delta_Z=0.1t$, we obtain $\lambda^{\rm crit}\simeq 0.94t$. This is
an interesting feature for applications since very strong Rashba
couplings are not necessary to generate the antiresonance.

\begin{figure}[t]
\centerline{
\epsfig{file=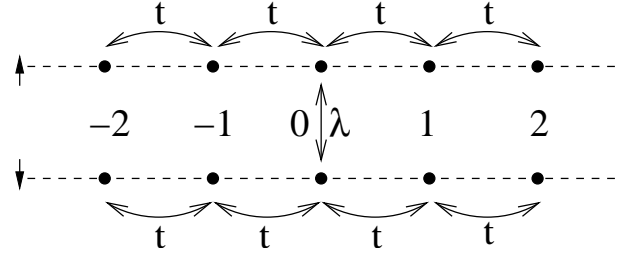,angle=0,width=0.45\textwidth,clip}
}
\caption{Sketch of the system considered in the discussion
for a point-like spin-flip interaction between propagating
(spin down) and evanescent (spin down) states along a
one-dimensional site lattice.}
\label{figtb3}
\end{figure}

The considerations made above suggest that the antiresonance
has a Fano lineshape $\propto (\epsilon+q)^2/(\epsilon^2+1)$
but it is hard to demonstrate with controlled approximations
that Eq.~(\ref{eq_tau}) has exactly this form.
Perhaps it is more instructive to consider a closely related model, in
which the spin flip interaction due to the Rashba coupling
is restricted to a single point, see Fig. \ref{figtb3}.
Then, the Hamiltonian reads
\begin{equation}
\label{eq_h3}
{\cal H}_{sf}=\sum_i s\frac{\Delta_Z}{2} c_{i\sigma}^\dagger c_{i\sigma}
-\sum_{\langle ij\rangle} t (c_{i\sigma}^\dagger c_{j\sigma}+{\rm H.c.})
+\lambda c_{0\uparrow}^\dagger c_{0\downarrow}+{\rm H.c.})\,.
\end{equation}
As before, the spin-up and spin-down
energy bands are shifted  by a Zeeman splitting.
The coupling between spins at the central site 
may represent the action of external magnetic
field pointing in a direction perpendicular to the spin quantization
axis or other source of spin flipping.
Spin flip interactions in quantum dots have recently received
much attention.\cite{sf1,sf2,sf3,sf4}

We take the wave function ansatz
$\psi_{n\downarrow}=e^{i k_\downarrow n}+re^{-i k_\downarrow n}$ for $n\le -1$,
and $\psi_{n\downarrow}=\tau e^{i k_\downarrow n}$ for $n\ge 1$
for the propagating states and
$\psi_{n\uparrow}=c e^{k_\uparrow n}$ for $n\le -1$ and
$\psi_{n\uparrow}=d e^{-k_\uparrow n}$ for $n\ge 1$
for the evanescent states. From the tight-binding equations,
we obtain the transmission amplitude,
\begin{equation}
\tau=\frac{2it\sin k_\downarrow a}{2it\sin k_\downarrow a+
\frac{\lambda^2}{2t\sinh k_\uparrow a}}
\,.
\end{equation}
We can define the broadening $\Gamma$ which measures the coupling
strength between spin-up and spin-down states and is proportional
to the density of states (per unit length) for electrons with spin down.
It reads,
\begin{equation}
\Gamma=\lambda^2\rho_\downarrow=\frac{\lambda^2}{2t\sin k_\downarrow a}\,.
\end{equation}
Using this result and the expressions for $k_\uparrow$ and $k_\downarrow$
written above, we find the expression for the transmission probability,
\begin{equation}\label{eq_tfano}
T=|\tau|^2=\frac{(E-\Delta_Z/2)^2-(2t)^2}
{(E-\Delta_Z/2)^2-(2t)^2+\Gamma^2} \,,
\end{equation}
valid for energies around the $T=0$ point $E=-2t+\Delta_Z/2$.
We note that Eq.~(\ref{eq_tfano}) has the desired Fano form.
In the conventional Fano effect, the coupling
takes place between a bound state immersed in a continuum band
and the propagating states.\cite{fano} Here, the role of the bound state
is played by the evanescent modes which, due to the spin flip
interaction, are coupled to the propagating states with opposite
spins.

\subsection{Angular dependence}\label{sec-1d-ang}
We now discuss the angular dependence of the $\vec{B}$ field direction.
In the strict 1D case, only one mode is needed. Thus, we expand
the wavefunction in the two-spinor basis which, in the case of an
in-plane field $\vec{B}=(B\cos\theta,B\sin\theta,0)$, takes the form:
\begin{equation}
\Psi(x)=\psi_1(x)\chi_+(\eta)+\psi_2(x)\chi_-(\eta)
\,,
\end{equation}
with
\begin{equation}\label{eq_chi}
\chi_\pm= \frac{1}{\sqrt{2}}\left( \begin{array}{ccc}
1 \\
\pm e^{i\theta}\end{array} \right)
\,,
\end{equation}
the spinors in the $\vec{B}$ direction.

\begin{figure}
\centerline{
\epsfig{file=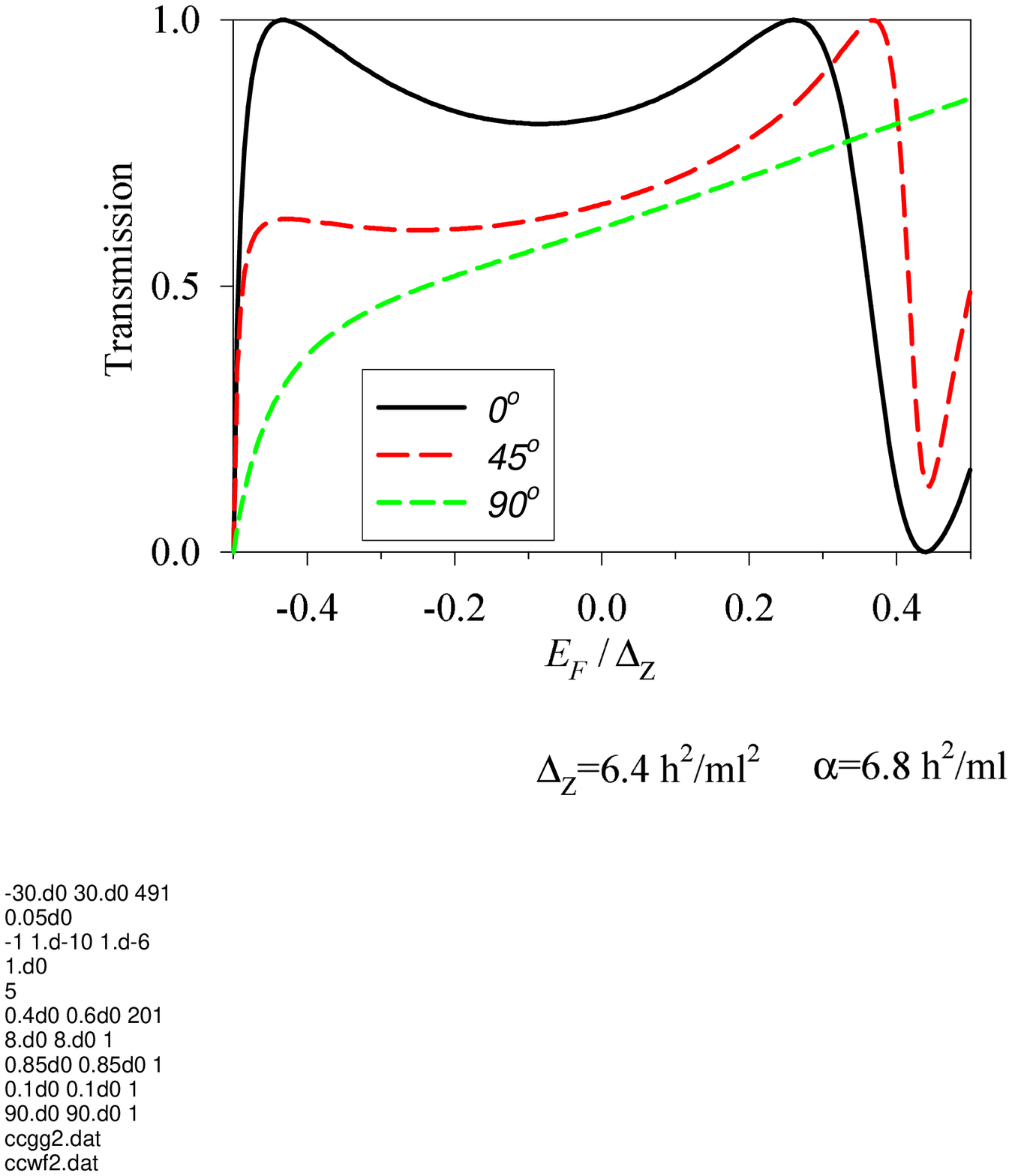,angle=0,width=0.4\textwidth,clip}
}
\caption{(Color online) Transmission as a function of Fermi energy
varying the orientation of the in-plane 
magnetic field. The azimuthal angle $\theta$ for each curve is given 
in the legend. 
We take $\alpha=6.8 \hbar^2/ml$ and $\Delta_Z=6.4\hbar^2/ml^2$.}
\label{theta}
\end{figure}

Substituting $\Psi$ in the Schr\"odinger equation with the Hamiltonian
given by Eq.~(\ref{ham1d}),
we obtain a pair of coupled equations for $\psi_1$ and $\psi_2$:
\begin{eqnarray}
\psi_1''&-&2ik_R\sin\theta\psi_1'+(k_F^2-k_B^2-ik_R'\sin\theta)\psi_1\nonumber\\
&=&-(2k_R\psi_2'+k_R'\psi_2)\cos\theta\label{eq_ccm1d1}\\
\psi_2''&+&2ik_R\sin\theta\psi_2'+(k_F^2+k_B^2+ik_R'\sin\theta)\psi_2\nonumber\\
&=&(2k_R\psi_1'+k_R'\psi_1)\cos\theta\label{eq_ccm1d2}
\,.
\end{eqnarray}
where primes indicate $d/dx$ and we recall $k_R(x)=m\alpha(x)/\hbar^2$.
We use the following gauge transformation,
\begin{eqnarray}\label{eq_psi12}
\psi_1&=&\tilde\psi_1 e^{i\sin\theta\xi(x)}\\
\psi_2&=&\tilde\psi_2 e^{-i\sin\theta\xi(x)}
\,,
\end{eqnarray}
where $\xi(x)=\int^x{dx'}k_R(x')$,
in order to eliminate the first derivatives in Eqs.~(\ref{eq_ccm1d1},\ref{eq_ccm1d2}),
which are transformed into
\begin{equation}\label{eq_hamccm1d}
\left( \begin{array}{cc}
H_{11} & H_{12}\\
H_{21} & H_{22}
\end{array} \right) 
\left( \begin{array}{c}
\tilde\psi_1 \\
\tilde\psi_2
\end{array} \right) =E_F
\left( \begin{array}{c}
\tilde\psi_1 \\
\tilde\psi_2
\end{array} \right)
\,,
\end{equation}
where the elements of the Hamiltonian matrix are
\begin{eqnarray}\label{eq_ccm1dm}
H_{11}&=&\lambda\left(\frac{d^2}{dx^2}+k_R^2\sin^2\theta-\frac{m\Delta_Z}{\hbar^2}\right) \\
H_{12}&=&\lambda\cos\theta e^{-2i\xi\sin\theta}
\left(2ik_R^2\sin\theta-k_R'-2k_R\frac{d}{dx}\right) \\
H_{21}&=&\lambda\cos\theta e^{2i\xi\sin\theta}
\left(2k_R\frac{d}{dx}+2ik_R^2\sin\theta+k_R'\right) \\
H_{22}&=&\lambda\left(\frac{d^2}{dx^2}+k_R^2\sin^2\theta+\frac{m\Delta_Z}{\hbar^2}\right) 
\,.
\end{eqnarray}
with $\lambda=-\hbar^2/2m$.
It is clear that for $\theta=\pi/2$ the amplitudes $\tilde\psi_1$ and $\tilde\psi_2$
decouple and for energies inside the gap the transmission is given simply by
the propagating state $\tilde\psi_2$.
Then, propagating and evanescent states become decoupled and no dip is expected.
Only for angles away from $\pi/2$ does the problem become
nontrivial since $\tilde\psi_2$ couples with the evanescent amplitude $\tilde\psi_1$.
Numerical results shown in Fig. \ref{theta} confirm this expectation.

\section{QUASI-ONE-DIMENSIONAL WIRE}\label{sec-q1d}

The preceding section has shown that the Fano resonance phenomenon 
manifests itself in 1D quantum wires with Zeeman splitting and a 
local Rashba interaction. Real wires, of course, have always 
some small extension in the lateral
direction. In this section we analyze the influence of 
this extra dimension by considering a quasi-1D wire, including 
the $y$ direction. This lateral $y$-confinement
is usually weaker than the vertical $z$-confinement. 
Thus, we neglect the contribution of the $y$-confining electric
field to the Rashba strength. 
For simplicity, we consider a transverse
potential of parabolic type, $m\omega_0^2y^2/2$.
\begin{figure}
\centerline{
\epsfig{file=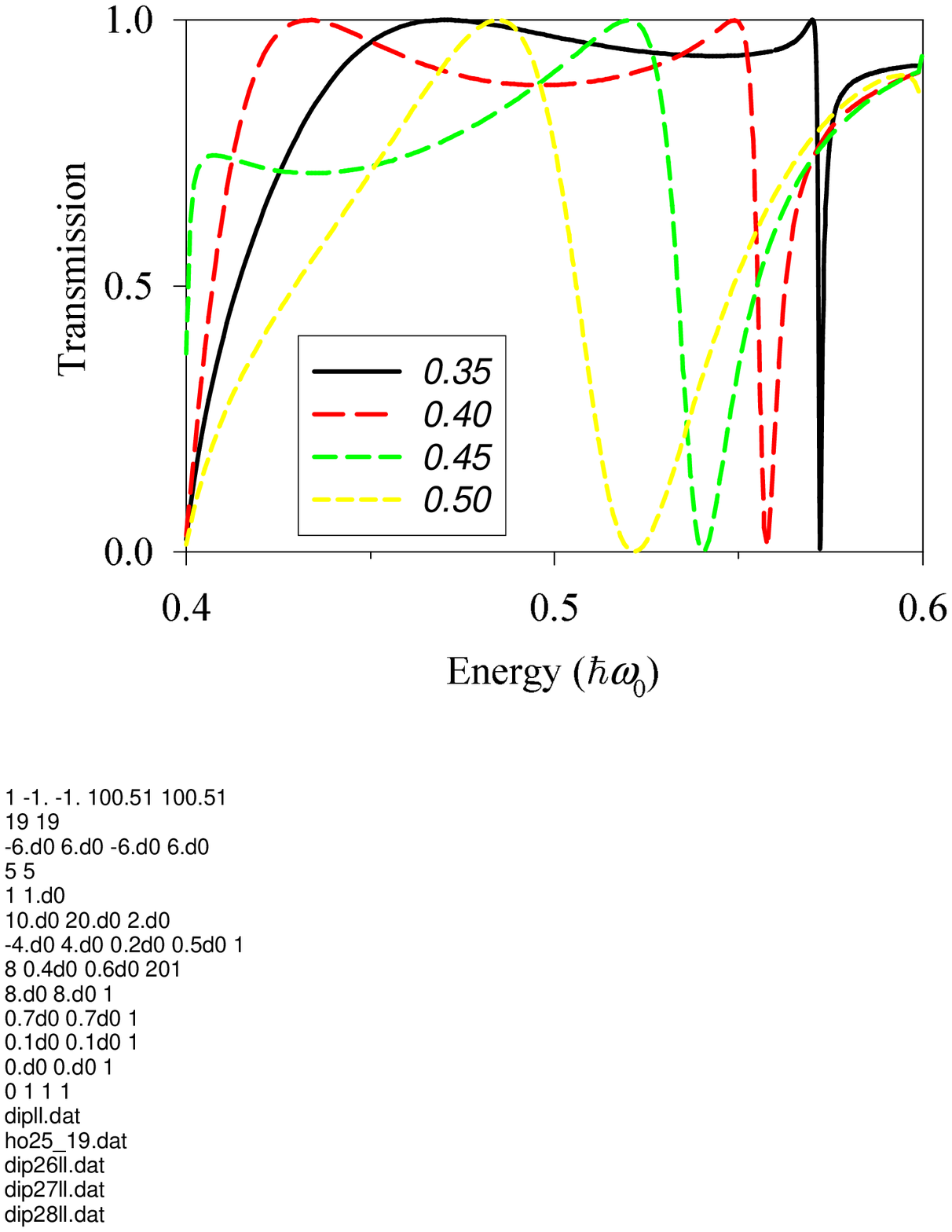,angle=0,width=0.4\textwidth,clip}
}
\caption{(color online) Transmission as a function of the Fermi energy for a quasi 1D wire
with transverse parabolic confinement characterized by $\omega_0$. 
A Rashba region of length $ l=8 l_0$ and 
a Zeeman energy $\Delta_Z=0.2\hbar\omega_0$ with the magnetic field along the wire 
($\theta=0$) have been used. 
The legend gives the numerical
values of $\alpha/\hbar\omega_0 l_0$ for the different curves.
}
\label{q1d1}
\end{figure}

The additional spatial
dimension is relevant now because the transverse momentum $p_y$ 
explicitly appears in the Rashba spin-orbit interaction as shown by
Eq.\ (\ref{Eq_Rash}).
It is also worth stressing that 
the new term in Eq.\ (\ref{Eq_Rash}), proportional to $p_y\sigma_x$,
precludes the use of
the analytical solution discussed in Sec.\ \ref{sec-1d-ext} for a wire with an extended 
Rashba interaction. In fact, it is well known that it causes the formation 
of textured spin states lacking well defined spin quantization axis 
even for a fixed value of the wavenumber $k$.\cite{gov02,ser05}
The Rashba coupling $\alpha(x)$ is assumed to be non zero only in a region of length $ l$,
where it takes the value $\alpha$,
as in Sec.\ \ref{sec-1d-di}. We also include the Zeeman coupling as in Sec.\ \ref{sec-1d-ext}
of a magnetic field oriented along a certain azimuthal angle $\theta$. The full
Hamiltonian thus reads
\begin{eqnarray}
\label{h2d}
{\cal H}_{\rm Q1D} &=&
\frac{p_x^2+p_y^2}{2m}+\frac{1}{2}m\omega_0^2y^2\nonumber\\
&+& \frac{\Delta_Z}{2}\left(\cos\theta\sigma_x+\sin\theta\sigma_y\right)
+{\cal H}_R\; .
\end{eqnarray}

A natural unit system for the present quasi-1D model is set by 
the wire transverse potential, with energy unit $\hbar\omega_0$ (oscillator energy) 
and length unit $ l_0=\sqrt{\hbar/m\omega_0}$ (oscillator length).
In what follows the numerical values for the Rashba region 
length $ l$, spin-orbit intensity $\alpha$
and Zeeman energy $\Delta_Z$ will be given in these oscillator units.
In order to obtain the transmission of the system 
modeled by Eq.\ (\ref{h2d})
we have used the 
quantum transmitting boundary algorithm as in Ref.\ \onlinecite{san06}.
The Schr\"odinger equation is discretized on a uniform grid using finite differences 
for the derivatives and imposing scattering boundary conditions. The reader is referred
to Refs.\ \onlinecite{san06,qtbm} for additional details of the method.
\begin{figure}
\centerline{
\epsfig{file=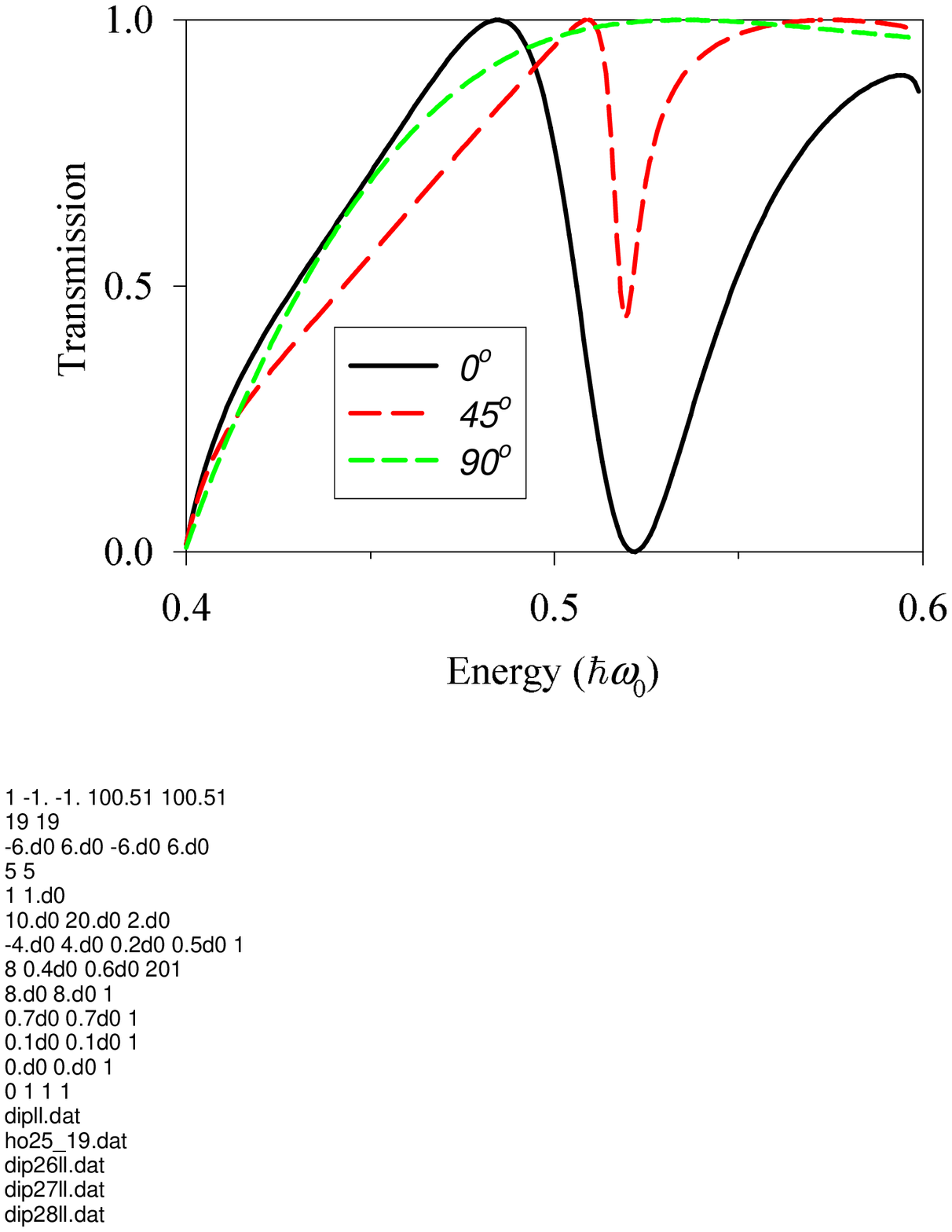,angle=0,width=0.4\textwidth,clip}
}
\caption{(Color online) Transmission versus Fermi energy varying the orientation of the in-plane 
magnetic field. The azimuthal angle $\theta$ for each curve is given 
in the legend. 
We use a Rashba region of length $ l=8 l_0$, 
Zeeman energy $\Delta_Z=0.2\hbar\omega_0$
and spin-orbit coupling strength $\alpha=0.5\hbar\omega_0 l_0$.}
\label{q1d3}
\end{figure}

The existence of the Fano lineshapes in a 
quasi-1D wire, with a transmission zero at a given energy, is clearly 
shown in Fig. \ref{q1d1}. This result proves that the physical effect
elucidated with the tight binding model of Sec.\ \ref{sec-1d-tb} is robust
and persists in more realistic models. 
There is also a nice qualitative agreement
with the 1D results of Fig.\ \ref{fig5}. In all three cases (tight-binding, 1D
and quasi-1D) increasing the value of
$\alpha$ leads to a shift towards lower energies of the transmission zero,
and to an important broadening of the transmission dip. 
These are very appealing features related to
practical applications in spintronic devices, since they 
could allow to control the transmission by tuning $\alpha$; the 
device operation would not be
very sensitive to small changes in $\alpha$ due to the broadness
of the dip.

The scales used in Fig. \ref{q1d1} are of the same order as in
Fig.\ \ref{fig5}. E.g., for a confinement strength $\hbar\omega_0=0.1$~meV
in an InAs wire, we obtain $l\simeq 1.5$~$\mu$m, $\Delta_Z\simeq 0.02$~meV
($B\simeq 20$~mT) and $\alpha\simeq 9$~meV~nm.

When the magnetic field is oriented along the wire as 
in Fig.\ \ref{q1d1}, the 
interference leading to the Fano profiles in the transmission is maximal.
On the contrary, for transverse orientation $\theta=\pi/2$ it completely 
disappears (see Fig.\ \ref{q1d3}). This behavior is in agreement with the analysis
of Sec.\ \ref{sec-1d-ang} in the 1D case, where it was shown that the mixings $H_{1,2}$ and 
$H_{2,1}$ vanish for $\theta=\pi/2$.

\begin{figure}
\centerline{
\epsfig{file=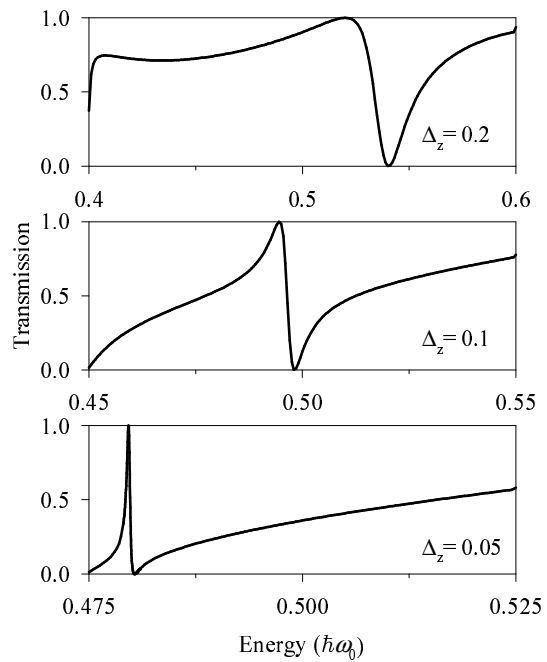,angle=0,width=0.4\textwidth,clip}
}
\caption{(Color online) Transmission as a function of the Fermi energy for a quasi 1D wire
with a Rashba region of $ l=8 l_0$ and a magnetic field along $\theta=0$.
The different panels correspond to  the given
Zeeman energies (in units of $\hbar\omega_0$).
In each panel solid and dashed lines correspond, respectively, to a value of 
$\alpha/\hbar\omega_0 l_0$ of 0.45 and 0.40, respectively.}
\label{q1d2}
\end{figure}

The evolution with the Zeeman field intensity for the quasi-1D case is shown
in Fig.\ \ref{q1d2}. The behavior is again qualitatively similar to the 
1D case of Fig.\ \ref{fig6}, with the dip evolving towards smaller energies when 
decreasing the value of $\Delta_Z$. We also notice that, as predicted by 
the tight binding result, even for 
quite small Zeeman energies there is a dip in the transmission.

\begin{figure}
\centerline{
\epsfig{file=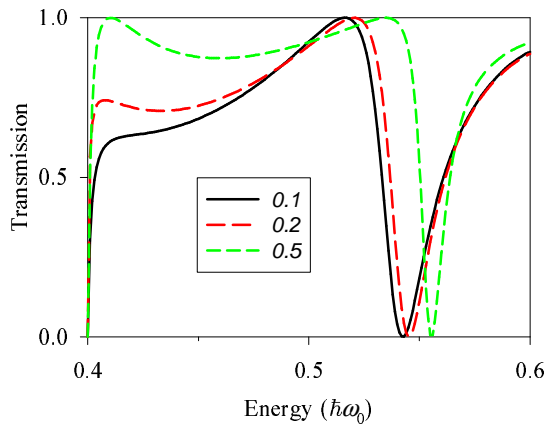,angle=0,width=0.4\textwidth,clip}
}
\caption{(Color online) Transmission versus Fermi energy
for $\alpha= 0.45\hbar\omega_0 l_0$ 
and $\Delta_Z=0.2\hbar\omega_0$. Each curve corresponds to a different diffusivity
$d$, as given in the legend in units of $ l_0$,
for a Rashba region of length $ l=8 l_0$}
\label{sig}
\end{figure}

Thus far we have considered abrupt interfaces between the normal sides
and the Rashba region. Using the quasi-1D grid calculation we
can also address the influence of the smoothness in the transition of the Rashba coupling 
strength from zero to the 
finite value $\alpha$, which is closer to reality. We model each interface 
using a Fermi function with a diffusivity $d$,
\begin{equation}
\alpha(x)=\alpha 
\left[
{1\over 1 + e^{(x- l/2)/d}}
-{1\over 1 + e^{(x+ l/2)/d}}
\right]\; .
\end{equation}
Figure \ref{sig} shows the results for different values of $d$.
The transmission curve coincides with the abrupt interface limit
when $d\sim 0.2 l_0$ while for increasing $d$ the interface
becomes smoother but the transmission zero remains visible
and the dip position does not change much.
This is a crucial observation---the transmission
zeros we find are roughly independent of the precise profile of the
Rashba strength. This robustness has an obvious importance for potential
spintronic applications.

\section{CONCLUSIONS}\label{sec-concl}
We have performed a theoretical analysis of the transport properties
of a ballistic quantum wire with spatially inhomogeneous Rashba interaction
in the presence of an external magnetic field giving rise to Zeeman
spin splitting. When the Rashba coupling dominates the magnetic field
an energy pseudogap develops in the wire spectrum. We find abrupt
transmission lineshapes when the Fermi energy lies within
the pseudogap. The lineshapes are narrow and asymmetric and the transmission
reaches zero for energies near the gap closing.
We have discussed a minimal tight-binding model that reproduces the
essential features of the resonances, yielding analytical expressions
for the lineshape dependence on Fermi energy, Rashba intensity and
Zeeman splitting. Qualitatively,
the evanescent band plays the role of a quasi-bound
state that the confined Rashba interaction couples to the
propagating states. The evanescent waves are not true bound states
but when the Fermi energy
approaches the evanescent band bottom electrons scattering off
the Rashba region become strongly affected, leading
to perfect reflection. Numerical results in realistic quantum
wires agree with the purely 1D case. Finally, we have analyzed
the behavior of the resonances when the angular orientation
of the magnetic field is changed and the interfaces become smoother.

The system studied here could work as a current modulator device.
For slight variations of the Fermi energy, which can be externally
controlled, we have shown that the transmission changes dramatically
between two limit values (1 and 0) across the antiresonance.
Our proposal has a number of differences compared to the Datta-Das
spin transistor.\cite{dat90} First, the latter device modulates the current
independently of the energy of the injected electrons since the phase difference
$\Delta\theta=2m\alpha l/\hbar^2$ that governs the spin precession
is independent of the wavevector. Then, small changes of $\alpha$ or $l$
strongly affects the working points of the transistor whereas in our case
these points are not very sensitive to small variations of the external
parameters such as the external magnetic field, the Rashba strength
or the interface diffusivity. Moreover, a 100$\%$ current modulation
is hard to achieve in the Datta-Das transistor, especially when intersubband coupling
is taken into account, whereas in our case the modulation is rather
abrupt and is preserved even when adjacent subbands are coupled, an effect
which is unavoidable in real quantum wires.
We have reported results for the lowest spin-split subband
but have checked that similar pronounced dips are seen
in higher subbands. Our results differ also with those of
Ref. \onlinecite{san06} since in that case the antiresonances
reached zero only after a fine tuning of the parameters.
Here, our only requirement is that the Fermi energy should lie
within the spectrum pseudogap.

As far as the discussion in 1D wires is concerned, the field
directions orthogonal to the Rashba field (along $y$, according to
the parameterization of the Hamiltonian we have employed) are equivalent.
However, in the quasi-1D case one of these directions (the one perpendicular
to both the Rashba field and the electron propagation) induce
orbital effects. Here, we have restricted ourselves to fields
giving rise only to Zeeman splittings since the study of orbital effects
in inhomogeneous systems requires knowledge of the evanescent
states when the magnetic field is applied perpendicular to the wire plane.
This is not a trivial task and it seems to be promising avenue of
future research. Reference \onlinecite{deb05}
finds important changes in the spectrum structure of a quantum wire
with uniform Rashba interaction and perpendicular magnetic fields.
However, our sharp antiresonances show up even in the presence of
rather small Zeeman splittings.
Therefore, we expect that the dips should be still visible even
when orbital effects are taken into account provided the magnetic field
length is much larger than the confinement length.

In our discussion, we have neglected electron-electron interaction
effects, which may lead to Luttinger liquid effects in 1D ballistic wires
when the interactions are screened like in a wire with electric-field induced
spin-orbit interactions.\cite{hau01,gri05}
When Zeeman splittings are present \cite{lee05,dev05} the transmission seems
to be altered by electron-electron interactions, although these works neglect
the intersubband coupling term of the Rashba interaction.
In fact, for ballistic wires without Rashba coupling but multiple populated
subbands a simple mean-field approach demonstrates\cite{but03} that Coulomb interactions
are crucial to understand rectification effects observed in
nanojunction rectifiers. On the other hand, single-particle effects
are shown\cite{lop07} to lead to Coulomb blockade antiresonances of the Fano form.
Hence, further work is needed to clarify the influence of Coulomb
interactions in the conductance of a quantum wire with Zeeman splitting and
a localized Rashba interaction.

\section*{ACKNOWLEDGMENTS}
This work was supported by the Spanish MEC Grant No.\ FIS2005-02796
and the ``Ram\'on y Cajal'' program.


\begin{thebibliography}{90}
\bibitem{dat90}
S.~Datta and B.~Das, Appl. Phys. Lett. {\bf 56}, 665 (1990).
\bibitem{ras60}
E.I~Rashba, Fiz. Tverd. Tela (Leningrad)  {\bf 2}, 1224 (1960).
[Sov. Phys. Solid State  {\bf 2}, 1109 (1960)].
\bibitem{byc84}
Y. Bychkov and E. I. Rashba, J. Phys. C {\bf 17}, 6039 (1984).
\bibitem{sat99}
Y. Sato, S. Gozu, T. Kikutani, and S. Yamada,
Physica B {\bf 272}, 114 (1999)
\bibitem{mor99}
A.V.~Moroz and C.H.W.~Barnes,
Phys. Rev. B {\bf 60}, 14272 (1999).
\bibitem{mir01}
F. Mireles and G. Kirczenow, Phys. Rev. B {\bf 64}, 024426 (2001).
\bibitem{kis01}
A.A. Kiselev and K.W. Kim,
Appl. Phys. Lett. {\bf 78}, 775 (2001).
\bibitem{mol01}
L.W. Molenkamp, G. Schmidt, and G.E.W. Bauer,
Phys. Rev. B {\bf 64}, 121202(R) (2001).
\bibitem{egu02}
J.C. Egues, G. Burkard, and D. Loss 
Phys. Rev. Lett. {\bf 89}, 176401 (2003);
J.C. Egues, G. Burkard, D. S. Saraga, J. Schliemann, D. Loss,
Phys. Rev. B {\bf 72}, 235326 (2005).
\bibitem{gov02}
M. Governale and U. Z\"ulicke, 
Phys. Rev. B {\bf 66}, 073311 (2002).
\bibitem{fev02}
G. Feve, W.D. Oliver, M. Aranzana, and Y. Yamamoto,
Phys. Rev. B  {\bf 66}, 155328 (2002).
\bibitem{bul02}
E.N. Bulgakov and A.F. Sadreev,
Phys. Rev. B {\bf 66}, 075331 (2002).
\bibitem{and03}
E.A. de Andrada e Silva and G.C.L. Rocca,
Phys. Rev. B {\bf 67}, 165318 (2003).
\bibitem{stre03}
P. Streda and P. Seba
Phys. Rev. Lett. {\bf 90}, 256601 (2003).
\bibitem{scha04}
Th. Sch\"apers, J.~Knobbe and V.A~Guzenko
Phys. Rev. B {\bf 69}, 235323 (2004).
\bibitem{per04}
Yu.V. Pershin, J.A. Nesteroff, and V. Privman,
Phys. Rev. B {\bf 69}, 121306 (2004).
\bibitem{nes04}
J.A. Nesteroff, Yu.V. Pershin, and V. Privman,
Phys. Rev. Lett. {\bf 93}, 126601 (2004)
\bibitem{cah04}
M. Cahay and S. Bandyopadhyay,
Phys. Rev. B {\bf 69}, 045303 (2004).
\bibitem{wan04}
X.F. Wang, Phys. Rev. B {\bf 69}, 035302 (2004).
\bibitem{she04}
I.A. Shelykh and N.G. Galkin, Phys. Rev. B {\bf 70}, 205328 (2004).
\bibitem{zha05}
F. Zhai and H.Q. Xu,
Phys. Rev. Lett. {\bf 94}, 246601 (2005).
\bibitem{kno05}
J. Knobbe and Th. Sch\"appers,
Phys. Rev. B {\bf 71}, 035311 (2005) 
\bibitem{per05}
R.G. Pereira and E. Miranda,
Phys. Rev. B {\bf 71}, 085318 (2005).
\bibitem{rom05}
C.L. Romano, S.E. Ulloa, and P.I. Tamborenea,
Phys. Rev. B {\bf 71}, 035336 (2005)
\bibitem{deb05}
S.~Debald and B. Kramer, Phys. Rev. B {\bf 71}, 115322 (2005).
\bibitem{ser05}
Ll. Serra, D. S\'anchez, and R. L\'opez,
Phys. Rev. B {\bf 72}, 235309 (2005).
\bibitem{zhan05}
L. Zhang, P. Brusheim, and H.Q. Xu, 
Phys.\ Rev.\ B {\bf 72}, 045347 (2005).
\bibitem{zhan06}
S. Zhang, R. Liang, E. Zhang, L. Zhang, and Y. Liu
Phys. Rev. B {\bf 73}, 155316 (2006) .
\bibitem{rey06}
A. Reynoso, G. Usaj, and C.A. Balseiro,
Phys. Rev. B {\bf 73}, 115342 (2006).
\bibitem{san06}
D. S\'anchez and Ll. Serra,
Phys. Rev. B {\bf 74}, 153313 (2006).
\bibitem{jeo06}
J.-S. Jeong and H.-W. Lee,
Phys. Rev. B {\bf 74}, 195311 (2006).
\bibitem{sha06}
Th. Schäpers, V. A. Guzenko, M.G. Pala, U. Z\"ulicke,
M. Governale, J. Knobbe, and H. Hardtdegen,
Phys. Rev. B {\bf 74}, 081301(R) (2006). 
\bibitem{zha06}
L. Zhang, F. Zhai, and H.Q. Xu,
Phys. Rev. B {\bf 74}, 195332 (2006).
\bibitem{per07}
C.A. Perroni, D. Bercioux, V. Marigliano Ramaglia, and V. Cataudella,
J. Phys.: Condens. Matter {\bf 19}, 186227 (2007).
\bibitem{nev07}
A.H. Nevidomskyy and K. Le Hur,
arxiv:cond-mat/0608340.
\bibitem{nit97}
J.~Nitta, T.~Akazaki, H.~Takayanagi, and T. Enoki,
Phys. Rev. Lett. {\bf 78}, 1335 (1997).
\bibitem{eng97}
G.~Engels, J.~Lange, Th. Sch\"apers, and H. L\"uth,
Phys. Rev. B {\bf 55}, R1958 (1997).
\bibitem{gru00}
D. Grundler, Phys. Rev. Lett. {\bf 84}, 6074 (2000).
\bibitem{wee88}
B.J. van Wees, H. van Houten, C.W.J. Beenakker, J.G. Williamson,
L.P. Kouwenhoven, D. van der Marel, and C.T. Foxon,
Phys. Rev. Lett. {\bf 60}, 848 (1988).
\bibitem{wha88}
D.A. Wharam, T.J. Thornton, R. Newbury, M. Pepper, H. Ritchie, and
G.A.C. Jones, J. Phys. C {\bf 21}, L209 (1988).
\bibitem{chu89}
C.S. Chu and R.S. Sorbello,
Phys. Rev. B {\bf 40}, 5941 (1989).
\bibitem{bag90}
P.F. Bagwell, Phys. Rev. B {\bf 41}, 10354 (1990).
\bibitem{fai90}
J. Faist, P. Gu\'eret, and H. Rothuizen,
Phys. Rev. B {\bf 42}, R3217 (1990).
\bibitem{tek91}
E. Tekman and S. Ciraci
Phys. Rev. B {\bf 43}, 7145 (1991).
\bibitem{gur93}
S.A. Gurvitz and Y.B. Levinson, Phys. Rev. B {\bf 47}, 10578 (1993).
\bibitem{noc94}
J.U. N\"ockel and A.D. Stone, Phys. Rev. B {\bf 50}, 17415 (1994).
\bibitem{fano}
U. Fano, Phys. Rev. {\bf 124}, 1866 (1961).
\bibitem{cer73}
F. Cerdeira, T.A. Fjeldly, and M. Cardona,
Phys. Rev. B {\bf 8}, 4734 (1973).
\bibitem{gor00}
J. G\"ores {\em et al.}, Phys. Rev. B {\bf 62}, 2188 (2000).
\bibitem{kob02}
K. Kobayashi, H. Aikawa, S. Katsumoto and Y. Iye,
Phys. Rev. Lett. {\bf 88}, 256806 (2002).
\bibitem{val04}
M. Val\'{\i}n-Rodr\'{\i}guez, A. Puente, and Ll. Serra,
Phys. Rev. B {\bf 69}, 085306 (2004).
\bibitem{cse04}
J. Cserti, A. Csord\'as, and U. Z\"ulicke,
Phys. Rev. B {\bf 70}, 233307 (2004).
\bibitem{lop07}
R. L\'opez, D. S\'anchez and Ll. Serra,
arxiv:cond-mat/0610515 (2006), to appear in Phys. Rev. B.
\bibitem{ser07}
Ll. Serra, D. S\'anchez, and R. L\'opez,
arXiv:0705.1506 (2007).
\bibitem{bul99}
E.N. Bulgakov, K.N. Pichugin, A.F. Sadreev, P. Streda, and P. Seba,
Phys. Rev. Lett. {\bf 83}, 376 (1999).
\bibitem{and89}
T. Ando, Phys. Rev. B {\bf 40}, 5325 (1989).
\bibitem{sf1}
W. Rudzinski and J. Barnas, Phys. Rev. B {\bf 64}, 085318 (2001).
\bibitem{sf2}
R. L\'opez and D. S\'anchez, Phys. Rev. Lett. {\bf 90}, 116602 (2003).
\bibitem{sf3}
M.-S. Choi, D. S\'anchez and R. L\'opez,
Phys. Rev. Lett. {\bf 92}, 056601 (2004);
\bibitem{sf4}
B. Dong, G.H. Ding, H.L. Cui, X.L. Lei,
Europhys. Lett. {\bf 69}, 424 (2005).
\bibitem{qtbm}
C.S. Lent and D.J. Kirkner,
J. Appl. Phys. {\bf 67}, 6353 (1990).
\bibitem{hau01}
W. H\"ausler, Phys. Rev. B {\bf 63}, 121310 (2001).
\bibitem{gri05}
V. Gritsev, G.I. Japaridze, M. Pletyukhov, and D. Baeriswyl,
Phys. Rev. Lett. {\bf 94}, 137207 (2005).
\bibitem{lee05}
H.C. Lee and S.-R.E. Yang,
Phys. Rev. B {\bf 72}, 245338  (2005).
\bibitem{dev05}
P. Devillard, A. Crepieux, K. I. Imura, and T. Martin,
Phys. Rev. B {\bf 72}, 041309(R) (2005).
\bibitem{but03}
M. B\"uttiker and D. S\'anchez,
Phys. Rev. Lett. {\bf 90}, 119701 (2003).
\bibitem{son98}
A.M. Song, A. Lorke, A. Kriele, and J.P. Kotthaus,
W. Wegscheider and M. Bichler,
Phys. Rev. Lett. {\bf 80}, 3831 (1998).

\end{thebibliography}
\end{document}